\documentclass[conference]{IEEEtran}
\IEEEoverridecommandlockouts

\usepackage{enumitem}
\usepackage{cite}
\usepackage{amsmath,amssymb,amsfonts}
\usepackage{algorithmic}
\usepackage{graphicx}
\usepackage{textcomp}
\usepackage{xcolor}
\usepackage{url}
\usepackage{comment}
\usepackage{multirow}
\usepackage{float}
\usepackage{array}
\usepackage{listings}
\usepackage{subfigure}
\def\BibTeX{{\rm B\kern-.05em{\sc i\kern-.025em b}\kern-.08em T\kern-.1667em\lower.7ex\hbox{E}\kern-.125emX}}
\begin{document}

\title{Characterizing and Understanding Energy Footprint and Efficiency of Small Language Model on Edges}

\author{\IEEEauthorblockN{Md Romyull Islam, Bobin Deng, Nobel Dhar, Tu N. Nguyen, Selena He, Yong Shi, Kun Suo}
\IEEEauthorblockA{
Department of Computer Science, Kennesaw State University, GA, USA\\
Email:
\{mislam22, ndhar\}@students.kennesaw.edu, \{bdeng2, tu.nguyen, she4, yshi5, ksuo\}@kennesaw.edu}
}

\maketitle

\begin{abstract}
Cloud-based large language models (LLMs) and their variants have significantly influenced real-world applications. Deploying smaller models (i.e., small language models (SLMs)) on edge devices offers additional advantages, such as reduced latency and independence from network connectivity. However, edge devices' limited computing resources and constrained energy budgets challenge efficient deployment. This study evaluates the power efficiency of five representative SLMs — Llama 3.2, Phi-3 Mini, TinyLlama, and Gemma 2 on Raspberry Pi 5, Jetson Nano, and Jetson Orin Nano (CPU and GPU configurations). Results show that Jetson Orin Nano with GPU acceleration achieves the highest energy-to-performance ratio, significantly outperforming CPU-based setups. Llama 3.2 provides the best balance of accuracy and power efficiency, while TinyLlama is well-suited for low-power environments at the cost of reduced accuracy.
In contrast, Phi-3 Mini consumes the most energy despite its high accuracy. In addition, GPU acceleration, memory bandwidth, and model architecture are key in optimizing inference energy efficiency. Our empirical analysis offers practical insights for AI, smart systems, and mobile ad-hoc platforms to leverage tradeoffs from accuracy, inference latency, and power efficiency in energy-constrained environments.
\end{abstract}

\begin{IEEEkeywords}
Edge Computing, Energy Efficiency, Small Language Models, Power Optimization, Artificial Intelligence
\end{IEEEkeywords}

\section{Introduction}
The growing complexity of machine learning models, particularly large language models (LLMs) and their variants, small language models (SLMs), has led to remarkable advancements in natural language processing (NLP) and reasoning tasks. However, this complexity has also resulted in significantly higher energy consumption, raising concerns about the sustainability of deploying such models, especially in resource-constrained edge environments~\cite{desislavov2021compute}. As models continue to evolve, inference has become a dominant factor in energy consumption, often surpassing training in real-world scenarios. For instance, a single ChatGPT query requires 2.9 watt-hours of electricity, compared with 0.3 watt-hours for a Google search, according to the International Energy Agency~\cite{ai-energy}. Edge devices like the Raspberry Pi 5 and Nvidia Jetson Orin are widely used for real-time AI applications, but their strict power and thermal constraints make energy-efficient inference a critical requirement.

The authors in~\cite{rungsuptaweekoon2017evaluating} show that energy efficiency varies significantly across different hardware platforms, underscoring the need for tailored optimization strategies. While advances in hardware have contributed to improved power efficiency, as illustrated in Figure~\ref{fig:gflop_per_wat}, which shows a steady increase in GFLOPS per watt over time, deploying SLMs on low-power devices remains a challenge. The increasing energy efficiency of GPUs has followed a nearly linear trend, but the computational demand of modern language models continues to grow at a faster rate, making it essential to explore model-specific optimizations for sustainable deployment. Beyond energy efficiency, cost considerations also play a significant role in AI deployment. Figure~\ref{fig:gflop_per_dollar} demonstrates the exponential growth in computing performance per dollar, indicating that hardware has become more cost-effective over time. However, reductions in hardware cost do not directly translate to lower energy consumption for SLM inference, particularly in edge environments where power constraints outweigh raw computational capability. This discrepancy further emphasizes the need to evaluate energy consumption patterns of various models across different platforms to inform better deployment strategies.

Recently, many model compression techniques, such as quantization, pruning, and knowledge distillation, have been widely explored to address these challenges. Quantization, in particular, reduces model size and power consumption by using lower bit-width representations, enabling efficient inference without significant accuracy loss \cite{wu2020integer}. However, there is still a lack of comprehensive studies that analyze the energy footprint and power consumption patterns of different quantized SLMs on various edge hardware platforms, particularly inference efficiency. Understanding these trade-offs is crucial for optimizing real-world deployments where energy constraints directly impact model usability.

The models were assessed using three widely recognized benchmarks: Massive Multitask Language Understanding (MMLU), which measures broad domain knowledge and reasoning ability; HellaSwag, which focuses on commonsense reasoning and contextual understanding; and Winogrande, which evaluates pronoun resolution and contextual reasoning~\cite{hendryckstest2021, zellers2019hellaswag, sakaguchi2021winogrande}. We used Otii Ace Pro to perform real-time energy measurements to assess model efficiency. This study examines multiple performance metrics, including throughput, energy per inference, accuracy per watt, and Energy-Delay Product (EDP), offering the first comprehensive analysis of how different SLMs behave under various constraints. This data contributes to optimizing SLM deployment strategies for power-efficient AI applications, helping researchers and practitioners select models that balance computational efficiency, inference speed, and energy constraints.

Based on our observation and analysis, we highlight our key contributions to improving energy-efficient AI deployment on edge platforms, as outlined below:
\begin{itemize}
    \item Detailed investigation of the energy efficiency and performance of representative SLMs on edges;
    \item Comprehensive power benchmarking, fine-grained energy measurements, and multiple perspective evaluation across most popular edge SLMs;
    \item Analysis of factors (i.e., GPU acceleration, memory bandwidth, model architecture, etc.) hindering the energy-effective deployment of SLM in edge platforms;
    \item Insights for energy-efficient deployment on edge devices and navigating the development and optimization of current and future edge SLMs.
\end{itemize}

The remainder of this paper is structured as follows. Section~\ref{sec:motivation} outlines  motivation for this study, discussing challenges of deploying SLMs in energy-constrained edge environments and the need for optimization strategies. Section~\ref{sec:related_work} reviews the relevant literature. Section~\ref{sec:Benchmark} describes the evaluation benchmarks used to assess model performance across multiple reasoning and contextual tasks. Section~\ref{sec:Methodology} details the methodology, including model selection, quantization techniques, and evaluation metrics for analyzing energy efficiency across various hardware platforms. Section~\ref{sec:Result} presents the results, providing an in-depth analysis of energy consumption, inference efficiency, and computational trade-offs across different models and devices. Finally, Section~\ref{sec:Conclusion} summarizes the key findings and outlines directions for future research.

\section{Background and Motivation}
\label{sec:motivation}

\begin{figure}
    \centering
    \includegraphics[width=0.93\linewidth]{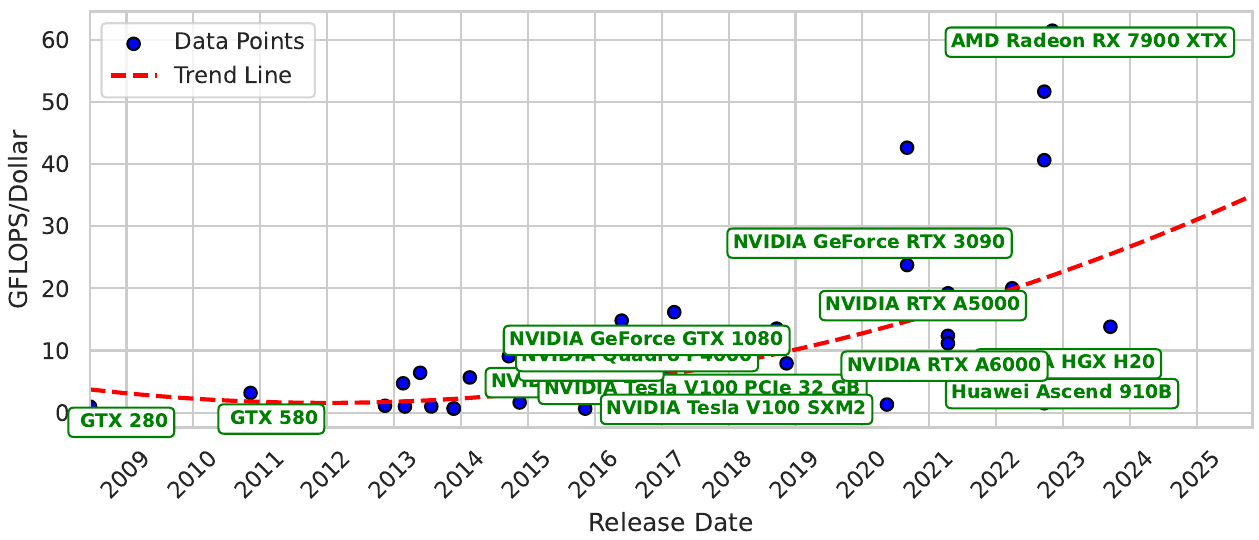}
    \vspace{-3mm}
    \caption{Full Precision (FP32) GPU Performance Trend over the Years Based on Hardware Price}
    \label{fig:gflop_per_dollar}
    \vspace{-4mm}
\end{figure}

\begin{figure}
    \centering
    \includegraphics[width=0.96\linewidth]{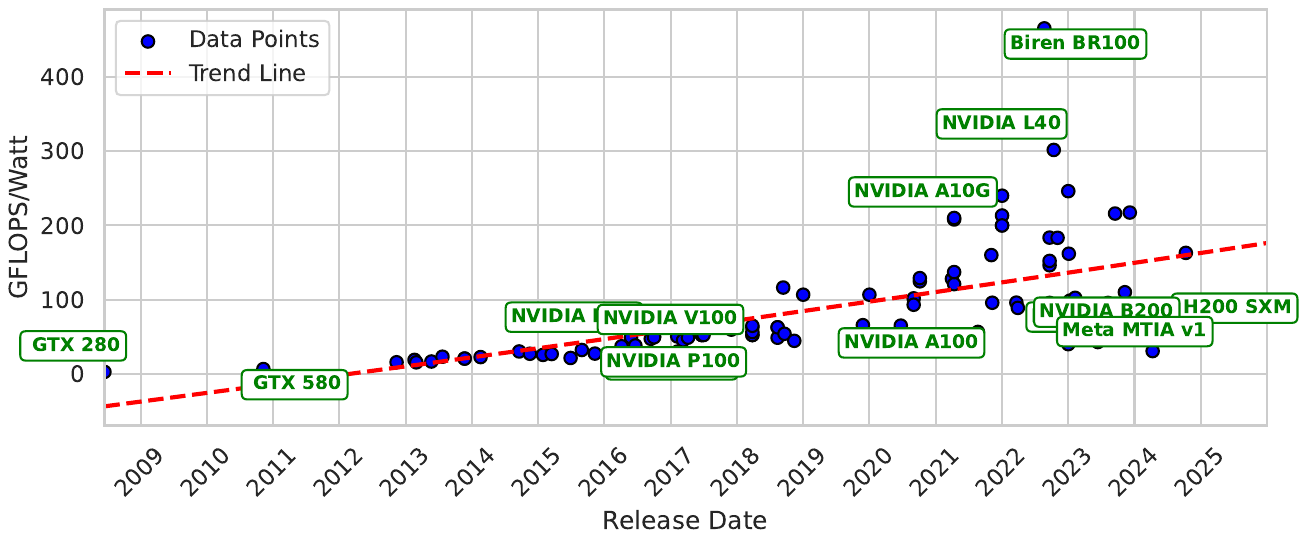}
    \vspace{-3mm}
    \caption{Full Precision (FP32) GPU Performance Trend over the Years Based on Energy Usage}
    \label{fig:gflop_per_wat}
    \vspace{-4mm}
\end{figure}

\textbf{Energy Efficiency in Edge Devices}. 
Unlike cloud environments, deploying language models on edge devices such as the Raspberry Pi 5 and Nvidia Jetson Orin requires careful power optimization due to strict energy and thermal constraints. While GPU computing costs per dollar have decreased by 30\% \cite{ml}, these improvements do not directly translate to efficient language model inference on resource-constrained systems. Figure~\ref{fig:gflop_per_dollar} illustrates the steady increase in GPU efficiency and the 1000 times reduction in AI inference cost per dollar over the last two decades. A similar trend is observed at the model level. For example, the cost per 1M tokens on GPT-4 dropped from \$36 in 2023 to \$4 on GPT-4o in 2024. However, as shown in Figure~\ref{fig:gflop_per_wat}, while GPU energy efficiency has also improved over the years, the rate of progress has been relatively modest. Deploying LLMs/SLMs on edge devices remains highly constrained by power consumption and energy footprint, necessitating targeted optimization strategies and further detailed investigation.

\vspace{3pt}
\textbf{Optimization of Inference Tasks}.
Language models exhibit varying energy consumption patterns across different hardware platforms, making it essential to analyze their inference efficiency systematically. Since power efficiency depends on the model architecture and deployment hardware, this study is interested in evaluating energy usage across the most popular SLMs on the most widely used edge platforms. The findings aim to identify power-performance trade-offs, recommending optimizations for energy-efficient inference.

\vspace{3pt}
\textbf{Real-time Measurement \& Understanding Trends}. 
Accurate energy profiling is crucial for understanding how different models behave on various hardware, especially on edge platforms. This research uses the latest Otii Ace Pro for real-time power monitoring, which enables precise energy consumption measurements during each inference. By analyzing EDP and related efficiency metrics, this research studies model-specific and hardware-dependent energy consumption patterns. The data collected provides insight into platform-specific energy trends on edges, helping to refine model selection and optimization strategies.

\begin{table*}[t]
\renewcommand{\arraystretch}{1.3}
\centering
\caption{Basic Specifications of Used Edge Devices}
\label{tab:device_spec}
\vspace{-1mm}
\begin{tabular}{cccccccccc}
\hline
Device Name & {Memory } & {Memory Freq} & {Memory Band} & {Memory Type}& {CPU Freq.} & {GPU Freq.} & {CPU Core} & {GPU Core}  & {Disk Size} \\ \hline
Raspberry Pi 5B & 4GB  & 4267MHz & 17GB/s & LPDDR4X & 2.4GHz & 0.0 & 4 & 0.0 & 128GB  \\
Jetson Nano & 4GB & 3200MHz & 25.6GB/s & LPDDR4 & 1.43GHz & 640MHz & 4 & 128 & 64GB  \\
Jetson Orin Nano & 8GB & 6375MHz & 102GB/s & LPDDR5 & 1.7GHz & 1020MHz & 6  & 1024 & 128 GB  \\ \hline
\end{tabular}
\vspace{-4mm}
\end{table*}

\section{Related Work}
\label{sec:related_work}

\subsection{Model-Side Optimizations}
Recent advancements in sparsity-based techniques and Low-Rank Adaptation (LoRA) have shown the possibility of improving the efficiency of LLM during inference. These methods, along with traditional model compression techniques such as quantization, achieve remarkable energy savings. For example, Argerich and Martínez \cite{argerich2024measuring} demonstrated that quantizing LLMs to 4-bit precision reduces memory usage by 50\%, maintaining accuracy while doubling inference speed on platforms like Raspberry Pi. Recent research has shown that combining quantization with advanced compression techniques can reduce model size while maintaining performance. For example, \texttt{Q4\_K\_M} quantization methods using k-means clustering have demonstrated significant efficiency gains \cite{paramanayakam2024less}. Structured pruning has also emerged as a key strategy, highlighted by Bartoldson et al.\cite{bartoldson2023compute}, who demonstrated that removing redundant weights or neurons significantly improves speed and energy efficiency. We selected \texttt{Q4\_K\_M} due to its superior trade-off between memory savings and accuracy retention, outperforming uniform 8-bit quantization in recent edge deployments~\cite{paramanayakam2024less}.

Furthermore, the exponential growth in model sizes has led to increased energy demands. Studies such as Fadel Argerich et al.\cite{argerich2024measuring} show how batch size adjustments, attention parallelization, and vocabulary reduction can optimize inference energy consumption. These techniques balance performance and computational cost, making LLMs more deployable in resource-constrained environments.

\vspace{-1mm}
\subsection{Hardware Efficiency and Device Trade-offs}
The choice of hardware platform plays a critical role in determining the energy consumption of LLM inference. Devices such as Nvidia Jetson Orin Nano, Jetson Nano, Raspberry Pi, and specialized AI accelerators each present unique power profiles and performance characteristics. A recent comparative study highlights that while Nvidia's Jetson series offers impressive processing capabilities, it requires significantly more power than lightweight devices like Raspberry Pi, which are better suited for simpler applications~\cite{better}. The Jetson Orin Nano delivers up to 40 TOPS of AI performance in a power-efficient form factor, making it particularly suitable for edge AI applications \cite{Solving}. Bekaroo et al.\cite{bekaroo2016power} demonstrated the energy efficiency of Raspberry Pi for lightweight computational tasks, emphasizing its suitability for power-sensitive environments.

In addition, devices with optimized power modes have been shown to provide substantial efficiency gains. Rungsuptaweekoon et al.\cite{rungsuptaweekoon2017evaluating} revealed that Jetson TX2's Max-Q mode achieves superior power efficiency compared to its Max-N mode, making it more suitable for real-time applications on embedded systems. Research has shown that different hardware platforms exhibit varying energy efficiency characteristics, with optimal performance dependent on specific deployment scenarios\cite{paramanayakam2024less}. These findings underscore the importance of selecting hardware that aligns with the application's energy and performance requirements.

Our work provides a comprehensive empirical comparison of energy consumption across multiple hardware platforms while evaluating the trade-offs between inference efficiency, memory constraints, and power consumption in real-world edge AI applications.

\subsection{Real-Time Energy Monitoring and Profiling}
Accurate measurement and monitoring of energy consumption are essential for optimizing LLM inference. Advanced monitoring frameworks like MELODI enable detailed analysis of energy consumption patterns during inference, providing crucial insights for optimization~\cite{husom2024price}. Systems such as the Otii Ace Pro, used by Müller et al.\cite{muller2024tinyep}, provide precise energy consumption data in edge environments, enabling developers to identify inefficiencies and optimize applications accordingly. Similarly, Rungsuptaweekoon et al.\cite{rungsuptaweekoon2017evaluating} employed the Low-Power Image Recognition Challenge (LPIRC) system to evaluate energy efficiency on embedded GPU platforms.

Integration with standardized benchmarks like MMLU and HellaSwag helps establish consistent evaluation metrics across different deployment scenarios~\cite{Benchmarks}. Recent studies~\cite{wang2024mmlu, samsi2023words} highlight how benchmarks can reveal patterns in energy usage across diverse tasks, guiding model optimization efforts. Additionally, tools like the Software Carbon Intensity (SCI) framework\cite{everman2023evaluating} have emerged as critical resources for assessing the environmental impact of LLM inference, emphasizing the need for energy-efficient deployment strategies.
Our work extends existing energy profiling frameworks by incorporating real-time power monitoring integrated with model-specific inference adjustments, enabling dynamic power adaptation and resource-aware execution strategies.

\section{Benchmark Description}
\label{sec:Benchmark}
The evaluation of our models was conducted using the MMLU benchmark \cite{hendryckstest2021}, HellaSwag \cite{zellers2019hellaswag}, and Winogrande \cite{sakaguchi2021winogrande}. These benchmarks were chosen to assess model performance across diverse tasks, including reasoning, knowledge retention, commonsense understanding, and contextual comprehension.

\subsubsection{MMLU Benchmark}
MMLU is a comprehensive benchmark comprising 57 tasks spanning various domains\cite{hendryckstest2021}. In our experiments, models were evaluated on ten subjects, including Abstract Algebra, Anatomy, Astronomy, Business Ethics, Clinical Knowledge, College Biology, College Chemistry, College Computer Science, College Mathematics, and College Medicine. Each task presents multiple-choice questions (A, B, C, D) requiring domain-specific knowledge and reasoning. By examining both computational performance and energy efficiency, MMLU provides a broad assessment of LLMs across different academic and professional fields.

\subsubsection{HellaSwag Benchmark}
HellaSwag is a multiple-choice dataset designed to evaluate models' ability to predict plausible following events in textual narratives, focusing on commonsense reasoning and contextual understanding\cite{zellers2019hellaswag}. Given a context from diverse sources such as activity descriptions, instructional texts, and storytelling, models must select the most logical continuation from four answer choices. The benchmark helps assess textual coherence, inference quality, and computational efficiency, complementing MMLU's broader knowledge-based evaluation.

\subsubsection{Winogrande Benchmark}
Winogrande is a multiple-choice dataset designed to assess commonsense reasoning~\cite{sakaguchi2021winogrande}. It builds upon the Winograd Schema Challenge, introducing a larger dataset and increased linguistic diversity to improve robustness. Each task presents a sentence with a blank space, requiring the model to select the most contextually appropriate option from two provided choices. By focusing on pronoun resolution and linguistic ambiguity, Winogrande delivers deeper insights into how models handle contextual dependencies and language understanding.

\section{Methodology}
\label{sec:Methodology}
This study evaluates the energy efficiency and performance of GGUF quantized language models deployed on edge devices. These models were selected for their diverse computational demands and architectural variations, ensuring a broad power efficiency comparison across different parameter sizes and inference capabilities. We used the edge devices listed in Table~\ref{tab:device_spec}. As device resource is constrained, we ensured all the model sizes remain within 4B parameters.
Llama 3.2 1B, a state-of-the-art lightweight LLM, is designed for efficient NLP tasks, balancing computational complexity and energy consumption. Phi-3 Mini is an optimized version of a high-capacity transformer, maintaining strong NLP capabilities with reduced power requirements, making it suitable for real-time applications. TinyLlama, a highly compact model, is tailored for deployment in hardware-constrained environments while demonstrating competitive accuracy on key NLP benchmarks. Gemma 2 prioritizes high-accuracy conversational AI while incorporating optimization techniques that improve inference speed without excessive energy draw. Table~\ref{tab:model_spec} provides further details on model sizes and training datasets.

\begin{table}[t]
\renewcommand{\arraystretch}{1.3}
\centering
\caption{Used Language Models and Their Parameters}
\vspace{-1mm}
\label{tab:model_spec}
\vspace{-1mm}
\begin{tabular}{ccc}
\hline
Model Name & Model Size & Tokens Trained on \\ \hline
TinyLlama & 1.1B & 3T \\
Phi-3 mini & 3.8B & 3.3T \\
Gemma 2 & 2B & 2T \\
Llama 3.2-1B & 1.24B & 9T \\ \hline
\end{tabular}
\vspace{-3mm}
\end{table}

\vspace{-1mm}
\subsection{Model Preparation}
Each model is downloaded from HuggingFace and is quantized to a 4-bit representation using k-means clustering-based Q4\_K\_M 4-bit GGUF format using \texttt{llama.cpp} framework. This technique clusters parameters based on their values before quantization, ensuring memory-efficient storage while reducing accuracy loss during inference. The quantized models are then deployed on various edge devices for evaluation.

\subsection{Data Acquisition}
The datasets used in this study were obtained from the Hugging Face Hub, covering a range of tasks to evaluate knowledge retention, commonsense reasoning, and contextual understanding. To balance evaluation diversity with practical inference time constraints on edge devices, we selected representative subsets from each dataset that capture the expected range of tasks. The test set of MMLU was picked to test broad knowledge and reasoning capabilities across several domains. 1,000 samples were extracted from various subjects, such as Abstract Algebra and College Medicine, ensuring a broad assessment of the model's ability to understand domain-specific information. For commonsense reasoning and logical inference, the HellaSwag validation dataset was utilized, focusing on 400 random validation tasks that required predicting the most plausible continuation of a given textual prompt. The Winogrande dataset, specifically its debiased evaluation set, was incorporated to examine pronoun disambiguation and contextual reasoning, using 500 samples to analyze how models handle linguistic ambiguity. By selecting these benchmarks, the study ensures a comprehensive evaluation of model efficiency across various reasoning tasks, offering insights into computational performance and energy trade-offs.

\vspace{-1mm}
\subsection{Evaluation Settings}
All models were evaluated in a zero-shot setting, meaning they were tested on tasks without prior fine-tuning or task-specific adaptation. This approach ensures that the results reflect the intrinsic generalization capabilities of each model. Metrics such as accuracy, latency, throughput, and energy consumption were analyzed to understand computational efficiency and practical deployment feasibility across different hardware platforms.

\vspace{-2mm}
\subsection{Performance Metrics}
\vspace{-0.5mm}
Performance is evaluated using the following metrics:

\textbf{Accuracy}: This metric reflects the percentage of correct answers provided by the models. Accuracy in the context of multiple choice questions can be represented as:
\begin{equation}
\text{Accuracy} = \left( \frac{\text{Number of Correct Predictions}}{\text{Total Predictions}} \right) \times 100
\end{equation}

\textbf{Latency}: The average time taken for each inference is measured, providing insights into the responsiveness of each model. Average latency can be calculated using the formula:
\begin{equation}
\text{Average Latency} = \frac{\text{Total Latency}}{\text{Total Number of Inferences}}
\end{equation}

\textbf{Throughput}: This is defined as the number of inferences completed per second, calculated as follows:
\begin{equation}
\text{Throughput} = \frac{\text{Total Number of Inferences}}{\text{Total Time Taken (in seconds)}}
\end{equation}

\textbf{Energy per Inference}: \textbf{Energy per Inference} Represents energy consumed per inference: 
\begin{equation} 
\text{Energy per Inference (Wh)} = \frac{\text{Energy (Wh)}}{\text{Total Inferences}} 
\end{equation} 
where energy (Wh) is the total energy consumed during the inference, and total inferences is the number of inference tasks.

\textbf{Energy-Delay Product (EDP)} is a key metric for evaluating system efficiency considering both energy consumption and delay\cite{vijaykumar2016framework}. It is calculated as follows:
\vspace{-2pt}
\begin{equation}
\text{EDP} = \text{Energy (J)} \times \text{Delay (s)}
\end{equation}
Where Energy (J) is derived from Energy (Wh) using:
\begin{equation}
\text{Energy (J)} = \text{Energy (Wh)} \times 3600
\end{equation}
EDP provides a holistic efficiency measure. A lower value indicates a system that is both faster and more energy-efficient.

\textbf{Energy-Delay Product per Billion Parameters (EDP/B)}: This metric captures the energy-delay efficiency scaled by model size:
\begin{equation}
\text{EDP/B} = \frac{\text{EDP (J·s)}}{\text{Model Size (Billion Parameters)}}
\end{equation}

\textbf{Watt-hours per Billion Parameters (Wh/B)}: This metric normalizes energy usage with respect to model size:
\begin{equation}
\text{Wh/B} = \frac{\text{Energy (Wh)}}{\text{Model Size (Billion Parameters)}}
\end{equation}

\subsection{Energy Monitoring}
Real-time voltage and current readings were collected using the Otii Ace Pro device to measure power consumption. Power (W) was computed as:
\begin{equation}
\text{Power (W)} = \text{Voltage (V)} \times \text{Current (A)}
\end{equation}
Energy usage over time was calculated as:
\begin{equation}
\text{Energy (Wh)} = \text{Power (W)} \times \text{Time (h)} 
\end{equation}
Measurements were recorded at 100Hz, allowing for fine-grained energy consumption tracking during inference. The power readings were synchronized with the inference timestamps, ensuring a precise correlation between performance and energy efficiency.

\subsection{Implementation Details}
\begin{figure}
    \centering
    \includegraphics[height=3.8cm]{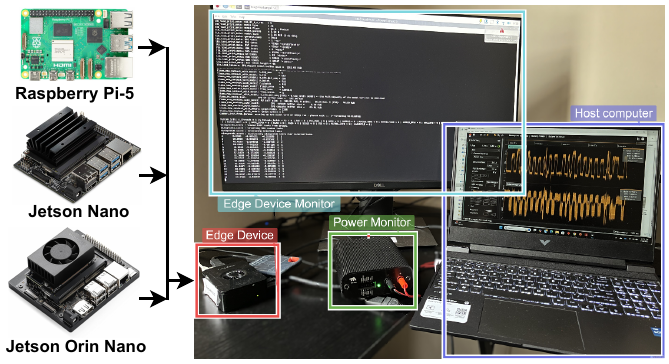}
    \vspace{-2mm}
    \caption{Setup of Power Measurement System}
    \label{fig:setup}
    \vspace{-4mm}
\end{figure}

The inference process was carried out using a Python script with \texttt{llama\_cpp\_python} library and \texttt{llama.cpp}, which ran the models, processed the datasets, and measured performance. To track power consumption, the Otii Ace Pro setup (Figure~\ref{fig:setup}) was used to continuously record energy usage, ensuring accurate measurements for each model. The power data was synchronized with the inference times, allowing a precise comparison of how different models consume energy under the same conditions.

\vspace{-1mm}
\section{Result Analysis}
\label{sec:Result}

\begin{table*}[ht]
\centering
\setlength{\tabcolsep}{4.5pt} 
\caption{Consolidated Performance Across Benchmarks and Devices}
\vspace{-0.8mm}
\label{tab:consolidated_performance}
\begin{tabular}{|c|c|ccc|ccc|ccc|ccc|}
\hline
\multirow{2}{*}{\textbf{Device}} & \multirow{2}{*}{\textbf{Model}} & \multicolumn{3}{c|}{\textbf{Acc. (\%)}} & \multicolumn{3}{c|}{\textbf{Energy (Wh)}} & \multicolumn{3}{c|}{\textbf{Latency (s)}} & \multicolumn{3}{c|}{\textbf{EDP (J·s)}} \\ \cline{3-14}
 & & \textbf{M} & \textbf{H} & \textbf{W} & \textbf{M} & \textbf{H} & \textbf{W} & \textbf{M} & \textbf{H} & \textbf{W} & \textbf{M} & \textbf{H} & \textbf{W} \\ \hline
\multirow{5}{*}{Raspberry Pi 5} & Llama 3.2 & 39.4 & 59.0 & 64.2 & 9.07 & 10.34 & 2.05 & 4.06 & 13.46 & 1.80 & 1.33e8 & 2.00e8 & 6.63e6 \\ 
                                & Phi-3 mini & 62.3 & 76.5 & 69.6 & 65.64 & 36.47 & 9.81 & 34.36 & 48.00 & 9.18 & 8.22e9 & 2.52e9 & 1.62e8 \\ 
                                & TinyLlama & 19.0 & 42.5 & 61.6 & 10.9 & 9.43 & 2.46 & 5.00 & 12.53 & 2.24 & 1.96e8 & 1.70e8 & 9.94e6 \\ 
                                & Gemma 2 & 30.4 & 68.25 & 69.0 & 32.77 & 23.15 & 5.36 & 17.17 & 30.55 & 5.05 & 2.03e9 & 1.02e9 & 4.87e7 \\ \hline
\multirow{5}{*}{Jetson Nano} & Llama 3.2 & 39.3 & 58.5 & 63.0 & 31.54 & 29.64 & 5.66 & 19.20 & 38.46 & 4.96 & 2.18e9 & 4.56e8 & 5.05e7 \\ 
                             & Phi-3 mini & \textbf{64.8} & 76.0 & 69.2 & 274.62 & 101.73 & 22.57 & 184.63 & 112.10 & 19.60 & \textbf{1.83e10} & \textbf{4.56e9} & \textbf{7.97e8} \\ 
                             & TinyLlama & 19.2 & 42.0 & 61.6 & 38.71 & 30.66 & 6.09 & 23.55 & 43.16 & 5.34 & 3.27e9 & 5.31e8 & 1.63e8 \\ 
                             & Gemma 2 & 33.8 & 67.5 & 69.0 & 67.76 & 67.65 & 12.14 & 41.23 & 94.23 & 10.80 & 6.78e9 & 2.55e9 & 2.36e8 \\ \hline
\multirow{5}{*}{Jetson Orin Nano} & Llama 3.2 & 39.8 & 58.5 & 63.4 & 9.71 & 5.83 & 1.24 & 3.92 & 5.47 & 0.92 & 1.37e8 & 1.28e7 & 5.67e5 \\ 
                                  & Phi-3 mini & 63.4 & 76.25 & 69.6 & 45.18 & 20.69 & 4.75 & 18.16 & 19.33 & 3.52 & 2.94e9 & 1.60e8 & 8.32e6 \\ 
                                  & TinyLlama & 18.0 & 41.75 & 62.4 & 12.04 & 5.59 & 1.28 & 4.90 & 5.24 & 0.94 & 2.12e8 & 1.17e7 & 6.17e5 \\ 
                                  & Gemma 2 & 33.6 & 67.75 & 68.8 & 20.75 & 12.88 & 2.75 & 8.39 & 12.52 & 2.05 & 4.99e8 & 6.44e7 & 2.82e6 \\ \hline
\multirow{5}{*}{Jetson Orin Nano (GPU)} & Llama 3.2 & 39.4 & 58.0 & 65.2 & 1.71 & 0.434 & 0.102 & 0.57 & 0.33 & 0.05 & 3.50e6 & \textbf{5.75e4} & 2.74e4 \\ 
                                  & Phi-3 mini & 64.3 & 76.25 & 70.2 & 6.56 & 1.05 & 0.305 & 2.43 & 1.04 & 0.18 & \textbf{5.74e7} & 4.36e4 & \textbf{2.79e5} \\ 
                                  & TinyLlama & 17.4 & 42.0 & 61.0 & 2.06 & 0.427 & 0.110 & 0.78 & 0.34 & 0.06 & 5.81e6 & 5.88e4 & 3.22e4 \\ 
                                  & Gemma 2 & 33.6 & 68.0 & 68.4 & 2.88 & 0.862 & 0.186 & 1.06 & 0.66 & 0.10 & 1.10e7 & 2.27e4 & 9.66e4 \\ \hline
\end{tabular}

\footnotesize{
\vspace{2mm}
Note: Accuracy (Acc.), Energy, Latency, and EDP are listed for MMLU (M), HellaSwag (H), and Winogrande (W). Latency is per inference.}
\vspace{-2mm}
\end{table*}

\begin{table*}[ht]
\centering
\caption{Throughput and Efficiency Metrics Across Benchmarks and Devices}
\label{tab:throughput_efficiency}
\vspace{-0.5mm}
\begin{tabular}{|c|c|ccc|ccc|ccc|}
\hline
\multirow{2}{*}{\textbf{Device}} & \multirow{2}{*}{\textbf{Model}} & \multicolumn{3}{c|}{\textbf{Ops/s \& Tokens/s}} & \multicolumn{3}{c|}{\textbf{Tokens/Wh}} & \multicolumn{3}{c|}{\textbf{Energy/Sec (W)}} \\ \cline{3-11}
 & & \textbf{M} & \textbf{H} & \textbf{W} & \textbf{M} & \textbf{H} & \textbf{W} & \textbf{M} & \textbf{H} & \textbf{W} \\ \hline
\multirow{5}{*}{Raspberry Pi 5} & Llama 3.2 & 0.25 & 12.21 & 16.46 & 110.25 & 6317.60 & 7187.32 & 0.00223 & 0.00192 & 0.00228 \\ 
                                & Phi-3 mini & 0.03 & 3.88 & 3.72 & 15.23 & 2041.87 & 1737.51 & 0.00189 & 0.00190 & 0.00214 \\ 
                                & TinyLlama & 0.20 & 14.99 & 15.30 & 91.74 & 7949.42 & 6969.51 & 0.00218 & 0.00188 & 0.00219 \\ 
                                & Gemma 2 & 0.06 & 5.34 & 5.75 & 30.52 & 2802.55 & 2701.68 & 0.00191 & 0.00190 & 0.00212 \\ \hline
\multirow{5}{*}{Jetson Nano} & Llama 3.2 & 0.05 & 4.26 & 5.95 & 31.71 & 2204.96 & 2603.18 & 0.00164 & 0.00193 & 0.00228 \\ 
                             & Phi-3 mini & 0.01 & 1.66 & 1.74 & 3.64 & 7322.72 & 755.44 & 0.00149 & 0.00227 & 0.00230 \\ 
                             & TinyLlama & 0.04 & 4.35 & 6.42 & 25.83 & 2445.04 & 2814.62 & 0.00164 & 0.00178 & 0.00228 \\ 
                             & Gemma 2 & 0.02 & 1.73 & 2.68 & 14.76 & 960.02 & 1193.77 & 0.00164 & 0.00180 & 0.00225 \\\hline
\multirow{5}{*}{Jetson Orin Nano} & Llama 3.2 & 0.26 & 30.05 & 32.24 & 102.98 & 11210.12 & 11801.61 & 0.00248 & 0.00266 & 0.00270 \\ 
                                  & Phi-3 mini & 0.06 & 9.64 & 9.70 & 22.13 & 3601.02 & 3588.42 & 0.00249 & 0.00268 & 0.00270 \\ 
                                  & TinyLlama & 0.20 & 35.84 & 36.38 & 83.02 & 13417.71 & 13316.41 & 0.00246 & 0.00267 & 0.00271 \\ 
                                  & Gemma 2 & 0.12 & 13.03 & 14.17 & 48.19 & 5042.06 & 5265.09 & 0.00247 & 0.00257 & 0.00268 \\ \hline
\multirow{5}{*}{Jetson Orin Nano (GPU)} & Llama 3.2 & 1.76 & 492.17 & 578.90 & \textbf{584.80} & 150647.00 & 144454.90 & 0.00301 & 0.00327 & 0.00379 \\ 
                                  & Phi-3 mini & 0.41 & 179.23 & 188.50 & 152.44 & 70954.29 & 55869.86 & 0.00269 & 0.00253 & 0.00333 \\ 
                                  & TinyLlama & 1.28 & 545.81 & 601.49 & 485.44 & \textbf{175690.16} & \textbf{155863.64} & 0.00263 & 0.00311 & 0.00377 \\ 
                                  & Gemma 2 & 0.94 & 246.81 & 296.46 & 347.22 & 75286.29 & 77854.84 & 0.00271 & 0.00328 & 0.00358 \\ \hline
\end{tabular}

\footnotesize{
\vspace{2mm}
Note: Ops/s(for MMLU) \& Tokens/s, Tokens/Wh, and Energy/Sec (W) are listed for MMLU (M), HellaSwag (H), and Winogrande (W).}
\vspace{-3mm}
\end{table*}

\subsection{Observations}

\subsubsection{Model Accuracy}
Model accuracy varied significantly across the MMLU, HellaSwag, and Winogrande benchmarks (Table~\ref{tab:consolidated_performance}). Phi-3 Mini consistently achieved the highest accuracy, scoring 62.3\% on MMLU, 76.5\% on HellaSwag, and 69.6\% on Winogrande. Llama 3.2 offered a balanced trade-off, maintaining solid accuracy (39.4\% MMLU, 58.5\% HellaSwag, 64.2\% Winogrande) with strong energy efficiency, making it suitable for resource-constrained deployments. Gemma 2 performed well in commonsense tasks but lagged in broader domains. TinyLlama had the lowest accuracy overall, highlighting its limitations in benchmarks that require reasoning.

\subsubsection{Energy Consumption}
Energy consumption varied significantly across devices, with Llama 3.2 consistently being the most power-efficient model (Table~\ref{tab:consolidated_performance}). On Raspberry Pi 5, it consumed as low as 2.05 Wh (Winogrande). In comparison, Jetson Orin Nano with GPU acceleration further reduced its power usage to 0.102 Wh (Winogrande), making it the best choice for energy-constrained edge deployments. Phi-3 Mini, in contrast, was the most energy-intensive model, requiring 65.64 Wh on Raspberry Pi 5 and peaking at 274.62 Wh on Jetson Nano, highlighting its inefficiency for edge applications. Even with GPU acceleration, it significantly reduced power consumption but remained the least efficient among all models. TinyLlama and Gemma 2 demonstrated moderate power consumption, with TinyLlama using 10.9 Wh on Raspberry Pi 5 and reducing to 2.06 Wh with GPU acceleration. Gemma 2 exhibited a similar trend, balancing power efficiency and performance. Llama 3.2 provided the best energy efficiency for edge AI applications.

\subsubsection{Latency}
Latency varied significantly across models and devices, with Phi-3 Mini consistently showing the highest inference times, making it the least suitable for real-time applications. On Raspberry Pi 5, it recorded 34.36 seconds for MMLU, while Jetson Nano amplified this issue, increasing MMLU latency to 184.63 seconds (Table~\ref{tab:consolidated_performance}). Llama 3.2 consistently showed the lowest latency, completing MMLU inference in 4.06 seconds on Raspberry Pi 5, 19.20 seconds on Jetson Nano, and 3.92 seconds on Jetson Orin Nano. GPU acceleration on Jetson Orin Nano further reduced latency to just 0.57 seconds, demonstrating its superior efficiency. TinyLlama had moderate latency, with TinyLlama recording 5.00 seconds for MMLU on Raspberry Pi 5, 23.55 seconds on Jetson Nano, and 4.90 seconds on Jetson Orin Nano.

\subsubsection{Throughput (Tokens/s \& Ops/s)}
Throughput analysis highlighted Llama 3.2 as the most consistent and efficient model across all benchmarks and devices (Table~\ref{tab:throughput_efficiency}). On Raspberry Pi 5, Llama 3.2 achieved the highest throughput, processing over 12 tokens per second (TPS) for HellaSwag, while TinyLlama followed closely with slightly lower TPS. Gemma 2 maintained moderate throughput, while Phi-3 Mini had the lowest performance, managing only a fraction of Llama 3.2. On Jetson Nano, Llama 3.2 and TinyLlama maintained high throughput, while Phi-3 Mini remained the least efficient. On Jetson Orin Nano, Llama 3.2 and TinyLlama significantly improved throughput, surpassing 30 TPS for HellaSwag and reaching even higher values with GPU acceleration. Llama 3.2 ultimately provided the most balanced performance across devices, demonstrating superior efficiency in handling inference tasks.

\subsubsection{Tokens per Watt-Hour (TPWh)}
TPWh analysis revealed significant efficiency variations between models and devices, with Llama 3.2 demonstrating the highest overall energy efficiency (Table~\ref{tab:throughput_efficiency}). On Raspberry Pi 5, Llama 3.2 achieved over 7,100 TPWh for Winogrande, outperforming TinyLlama, which excelled in HellaSwag but showed lower efficiency in other benchmarks. Gemma 2 followed closely, while Phi-3 Mini remained the least efficient, managing only about 2,000 TPWh for HellaSwag. On Jetson Nano, Llama 3.2 maintained its lead with more than 2,600 TPWh, while TinyLlama showed competitive efficiency. Jetson Orin Nano with GPU acceleration demonstrated the most striking efficiency gains, where Llama 3.2 achieved over 144,400 TPWh for Winogrande and TinyLlama peaked at over 175,600 TPWh for HellaSwag. Although TinyLlama excelled in specific cases, Llama 3.2 remained the most balanced model on all benchmarks. Gemma 2 delivered moderate efficiency, while Phi-3 Mini remained the least energy-efficient model even with GPU acceleration.

\subsubsection{Energy-Delay Product (EDP)}
EDP analysis confirmed Llama 3.2 is the most energy-efficient model across all devices and benchmarks (Table~\ref{tab:consolidated_performance}). On Raspberry Pi 5, it consistently stayed below $2.00 \times 10^8$~J$\cdot$s, followed by TinyLlama. Gemma 2 showed moderate efficiency, while Phi-3 Mini was the least efficient, reaching over $8.22 \times 10^9$~J$\cdot$s for MMLU. A similar trend appeared on Jetson Nano, where Phi-3 Mini exceeded $1.83 \times 10^{10}$~J$\cdot$s, reinforcing its inefficiency in constrained environments. On Jetson Orin Nano (GPU), EDP dropped significantly across all models. Llama 3.2 again led, followed by TinyLlama, while Phi-3 Mini remained the least efficient.

\begin{table*}[ht]
\centering
\caption{Normalized Energy and Efficiency Metrics per Billion Parameters Across Devices}
\label{tab:model_size_energy}
\begin{tabular}{|c|c|cc|cc|cc|cc|}
\hline
\textbf{Model} & \textbf{Size (B)} & 
\multicolumn{2}{c|}{\textbf{Raspberry Pi 5}} & 
\multicolumn{2}{c|}{\textbf{Jetson Nano}} & 
\multicolumn{2}{c|}{\textbf{Jetson Orin Nano}} & 
\multicolumn{2}{c|}{\textbf{Jetson Orin Nano (GPU)}} \\ \cline{3-10}
 & & \textbf{Wh/B} & \textbf{EDP/B (J·s/B)} & 
     \textbf{Wh/B} & \textbf{EDP/B (J·s/B)} & 
     \textbf{Wh/B} & \textbf{EDP/B (J·s/B)} & 
     \textbf{Wh/B} & \textbf{EDP/B (J·s/B)} \\ \hline
TinyLlama & 1.1 & 6.91 & 1.46e8 & 22.86 & 4.96e8 & 5.73 & 1.15e7 & 0.79 & 2.94e4 \\
Llama 3.2 & 1.24 & 5.77 & 1.16e8 & 17.97 & 3.04e8 & 4.51 & 9.58e6 & \textbf{0.60} & \textbf{2.42e4} \\
Gemma 2   & 2.0 & 10.22 & 2.15e8 & 24.59 & \textbf{4.83e8} & 6.06 & 1.17e7 & 0.66 & 4.43e4 \\
Phi-3 mini & 3.8 & 9.82 & 2.16e8 & \textbf{35.00} & 4.82e8 & 6.20 & 1.64e7 & 0.69 & 7.34e4 \\ \hline
\end{tabular}

\footnotesize{
\vspace{2mm}
Note: Wh/B = Watt-hours per Billion Parameters, EDP/B = Energy-Delay Product per Billion Parameters. Values are averaged across the benchmarks.}
\vspace{-3.5mm}
\end{table*}

\subsubsection{Energy per Sec}
This metric provided insights into the real-time power consumption of models across different benchmarks and devices (Table~\ref{tab:throughput_efficiency}). Phi-3 Mini consistently exhibited the highest power draw, peaks over 0.00230 W in Jetson Nano, emphasizing its inefficiency for edge deployments. In contrast, Llama 3.2 demonstrated the lowest overall power consumption, averaging around 0.002 W across devices, with GPU acceleration on Jetson Orin Nano further optimizing its efficiency. TinyLlama and Gemma 2 maintained moderate power usage, with TinyLlama slightly more efficient than Gemma 2 in most cases. They consumed less power than Phi-3 Mini but remained higher than Llama 3.2, following a similar trend across benchmarks.

\subsubsection{Accuracy Per Watt-Hour}
The accuracy per watt-hour metric evaluates the energy efficiency of models on different benchmarks (Figure~\ref{fig:accuracy_per_watt}). Llama 3.2 consistently demonstrated the highest efficiency, achieving over four accuracy points per watt-hour on Raspberry Pi and more than 23 on Jetson Orin Nano (GPU) for MMLU. TinyLlama followed with competitive efficiency, while Gemma 2 performed moderately. Phi-3 Mini remained the least efficient, with its performance falling below 0.3 accuracy points per watt-hour on Jetson Nano. A similar trend was observed for HellaSwag, where Llama 3.2 led with over 130 accuracy points per watt hour on Jetson Orin Nano (GPU), followed by TinyLlama. Gemma 2 showed lower efficiency, while Phi-3 Mini remained the least energy-efficient. For Winogrande, Llama 3.2 again maintained its lead, exceeding 600 accuracy points per watt-hour, with TinyLlama following closely. Gemma 2 performed moderately, while Phi-3 Mini continued to rank lowest across all benchmarks.

\subsubsection{Energy Per Inference}
\begin{figure*}
    \centering
    \includegraphics[width=0.99\linewidth]{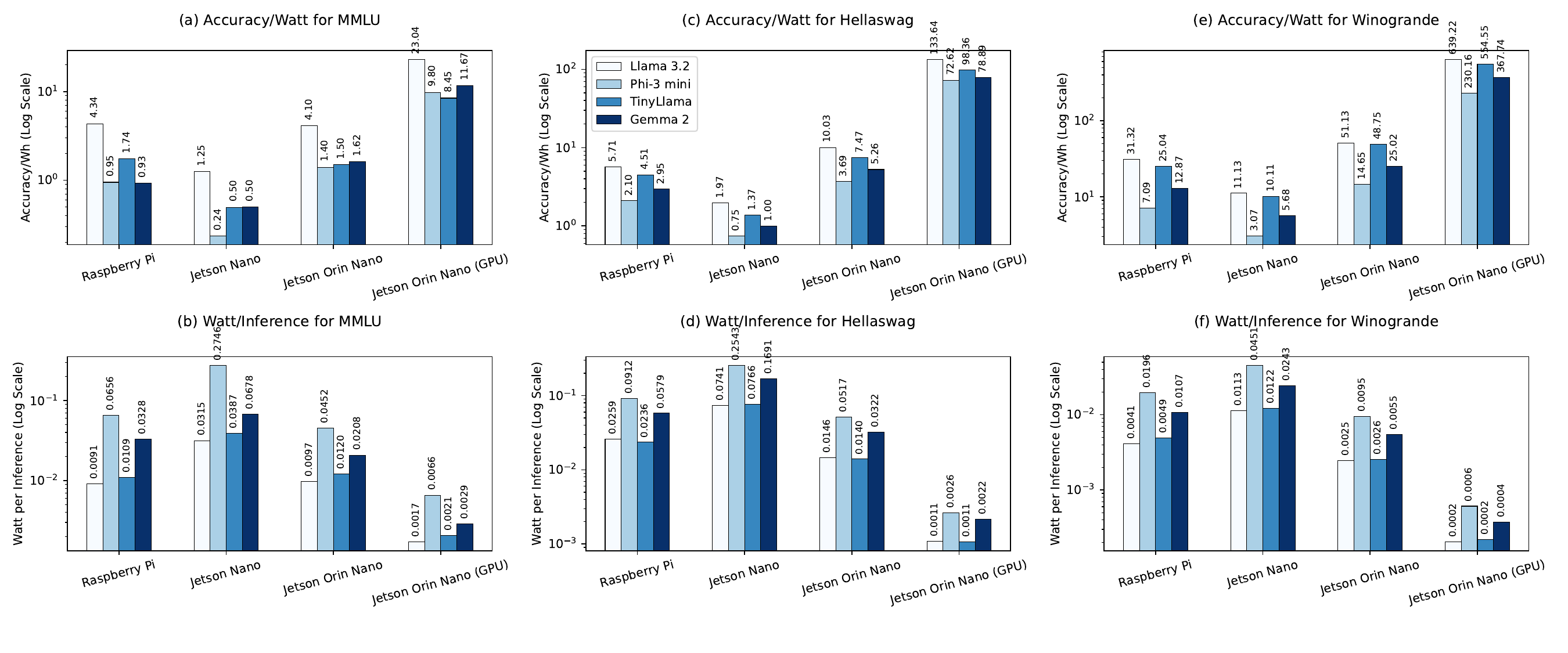}
    \vspace{-8mm}
    \caption{Prediction Accuracy Per Watt-Hour and Energy Consumption Per Inference Across Devices}
    \vspace{-1mm}
    \label{fig:accuracy_per_watt}
\end{figure*}

This metric quantifies the power required for each model to process a single inference, with lower values indicating better efficiency. Across all benchmarks, Llama 3.2 exhibited the lowest power consumption, requiring as little as 0.00171 Wh per inference on Jetson Orin Nano (GPU) for MMLU (Figure~\ref{fig:accuracy_per_watt}d). TinyLlama followed closely, demonstrating strong energy efficiency, while Gemma 2 consumed slightly more power. Phi-3 Mini remained the least efficient, requiring the most energy per inference across all benchmarks. This tendency remained consistent in HellaSwag and Winogrande, where Llama 3.2 maintained the lowest energy usage, followed by TinyLlama. Phi-3 Mini continued to rank lowest, consuming significantly more energy per inference than other models (Figure~\ref{fig:accuracy_per_watt}e,~\ref{fig:accuracy_per_watt}f).

\subsubsection{Runtime-Efficiency}
\begin{figure*}
    \centering
    \includegraphics[width=0.95\linewidth]{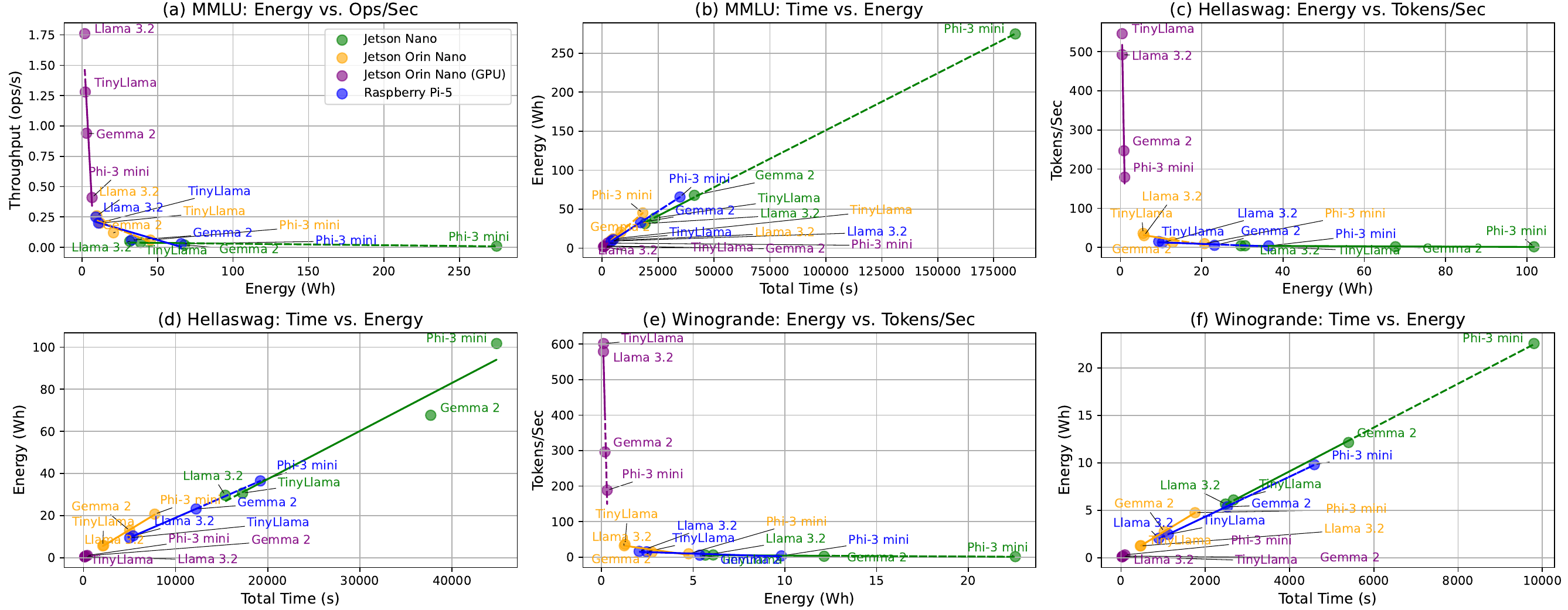}
    \vspace{-3.5mm}
    \caption{Comparison of Total Time, Energy Consumption, and Tokens Per Second for Benchmark Tasks Across Edge Devices}
    \vspace{-3.5mm}
    \label{fig:time_energy}
\end{figure*}

Deploying SLMs on edge devices requires balancing accuracy, energy consumption, latency, and throughput. Jetson Orin Nano (GPU) consistently outperformed CPU-based setups, reducing inference time and energy draw while maximizing throughput across all benchmarks (Figure~\ref{fig:time_energy}).
Llama 3.2 offered the best trade-off, combining high accuracy per watt-hour, low energy per inference, and strong throughput. It's 1.24B parameter size scaled well across both CPU and GPU. In contrast, the Phi-3 Mini showed the worst efficiency, particularly on Jetson Nano, where its 3.8B size led to long latency, high energy usage, and poor throughput even with GPU acceleration.

TinyLlama consumed among the least energy per inference and delivered high throughput per watt, especially on GPU. However, Llama 3.2 remained more efficient when both energy and accuracy were considered. Gemma 2 delivered moderate performance, with a better CPU-to-memory bandwidth ratio than Phi-3 Mini, contributing to improved efficiency.

\subsubsection{Normalized Energy Efficiency}
To fairly compare models of different sizes, we evaluate both energy usage (Wh/B) and Energy-Delay Product (EDP/B) per billion parameters, as summarized in Table~\ref{tab:model_size_energy}. These metrics capture not only the raw power draw but also the computational cost relative to model complexity.
Llama 3.2 consistently exhibited the best normalized efficiency across all platforms, achieving the lowest EDP/B of $2.42 \times 10^4$ J·s/B and Wh/B of 0.60 on Jetson Orin Nano (GPU), while staying under 6 Wh/B on all CPU-based setups. TinyLlama followed closely, particularly in low-power scenarios, reaching 0.79 Wh/B and $2.94 \times 10^4$ J·s/B on Jetson Orin Nano (GPU).

In contrast, Phi-3 Mini had the highest normalized energy cost, with 35.00 Wh/B and $4.82 \times 10^8$ J·s/B on Jetson Nano, highlighting the inefficiency of larger models on constrained hardware even with GPU support. Gemma 2 showed moderate energy-to-performance scaling, placing between TinyLlama and Phi-3 Mini in both metrics.
These findings reinforce that compact yet well-optimized models like Llama 3.2 and TinyLlama provide the most efficient trade-off between performance and energy use in edge environments, particularly when GPU acceleration is available.

\vspace{-1mm}

\subsection{Insights}
\label{sec:Insights}

\subsubsection{Power-Performance Trade-offs}
Achieving optimal power-performance trade-offs on edge devices requires balancing throughput, latency, and energy consumption within tight hardware constraints. Our results show that while GPU acceleration significantly boosts throughput and reduces inference time, it is not the sole factor in achieving energy efficiency.
Jetson Orin Nano (GPU) demonstrated the best trade-off across all benchmarks, combining high performance with low energy use. However, even the CPU-only Orin Nano outperformed Raspberry Pi 5 and Jetson Nano, highlighting the importance of CPU design, memory bandwidth, and system-level efficiency. These hardware characteristics directly impact how well a model scales in terms of power draw per unit of performance.

Smaller models like Llama 3.2 and TinyLlama benefited most from these optimized platforms, maintaining strong throughput with minimal energy cost. In contrast, Phi-3 Mini's high accuracy came at a steep energy cost, especially on CPU-limited devices. This underscores that the most performant models are not always the most power-efficient—effective deployment requires matching model size and hardware capabilities to the intended use case.

\subsubsection{Energy Efficiency Insights}
Energy efficiency in edge inference depends on accuracy per watt-hour, energy per inference, and hardware support. Jetson Orin Nano (GPU) consistently delivered the lowest energy per inference, demonstrating the value of GPU acceleration and high memory bandwidth in reducing power draw.
Among models, Llama 3.2 consistently ranked highest in energy efficiency across all benchmarks, balancing accuracy and energy cost-effectively (Figure~\ref{fig:accuracy_per_watt}a,~\ref{fig:accuracy_per_watt}d). TinyLlama achieved maximum token efficiency in HellaSwag, making it suitable for high-throughput, low-power scenarios. However, performance dropped on more complex benchmarks such as MMLU and Winogrande (Figure~\ref{fig:accuracy_per_watt}c,~\ref{fig:accuracy_per_watt}f). 

In contrast, Phi-3 Mini consumed the most energy, primarily due to its larger parameter size and memory demands, which limited its suitability for constrained environments. Gemma 2 showed moderate efficiency, benefiting from improved CPU and memory balance over Jetson Nano.
Normalized by model size, Llama 3.2 maintained the lowest energy-per-billion-parameters across all devices, reinforcing its suitability for sustained, efficient edge deployment.

\vspace{-0.02mm}
\subsubsection{Device Configuration and GPU Integration for Energy Optimization}
Hardware plays a vital role in energy-efficient edge AI inference. Jetson Orin Nano (GPU) consistently outperforms CPU-based setups due to its higher core count, parallel processing capabilities, and superior memory bandwidth (102GB/s). These advantages minimized inference latency and reduced power consumption, making GPU-accelerated execution significantly more efficient. The results emphasize that integrating a small, power-efficient GPU into edge devices drastically improves throughput and energy efficiency. Jetson Orin Nano (GPU) accelerated inference and reduced energy per operation, demonstrating the importance of optimized GPU pipelines in large-scale inference with minimal power overhead. In contrast, Jetson Nano exhibited the worst energy efficiency, as lower CPU clock speeds, limited memory bandwidth, and lack of parallelization increased both latency and energy consumption. CPU-bound inference struggled with larger models, reinforcing the need for dedicated accelerators for low-power AI applications. These findings confirm that hardware-aware optimizations, such as quantization, pruning, and mixed-precision computation, are essential for maximizing AI efficiency at the edge.

\vspace{-1.5mm}
\subsubsection{Recommendations for Edge Deployment}
Selecting the right model for edge AI requires balancing accuracy, power efficiency, and inference speed. Llama 3.2 offers the best trade-off, achieving high accuracy with minimal energy consumption, especially when paired with lightweight GPU acceleration. TinyLlama is optimal for ultra-low-power scenarios like wearables and mobile AI, while Phi-3 Mini, despite its superior accuracy, is impractical for energy-constrained applications due to high power demands. Gemma 2 provides a balanced alternative for general-purpose tasks where moderate efficiency is acceptable without GPU acceleration. GPU acceleration significantly improves inference speed and energy efficiency, as shown by Jetson Orin Nano. Devices with higher memory bandwidth and optimized GPU pipelines enable efficient LLM/SLM deployment, making them essential to sustainable edge AI applications.

For real-time applications requiring low latency (i.e., autonomous systems or interactive IoTs), Llama 3.2 on Jetson Orin Nano (GPU) is recommended due to its low latency (e.g., 0.57s for MMLU) and high throughput (e.g., 578.90 tokens/s for Winogrande). In contrast, for battery-powered devices where energy efficiency is paramount, TinyLlama on Raspberry Pi 5 or Jetson Orin Nano offers the best energy per inference (e.g., 0.110 Wh for Winogrande on Jetson Orin Nano GPU). Developers should also consider hybrid approaches, such as offloading initial model loading to the cloud while performing inference locally, to mitigate memory constraints on edge devices. Additionally, leveraging dynamic power management techniques, such as adjusting clock speeds based on workload, can further optimize energy usage during inference. These findings especially apply to autonomous mobile agents and IoT nodes in Mobile Ad-Hoc Networks, where model selection must accommodate intermittent connectivity, limited power, and latency-sensitive tasks.

\vspace{-0.5mm}
\section{Conclusion}
\label{sec:Conclusion}
This study benchmarks four quantized small language models (SLMs) across diverse edge hardware platforms, evaluating their energy efficiency and performance trade-offs. Llama 3.2 emerged as the most energy-efficient model, offering a strong balance between accuracy and power usage across MMLU, HellaSwag, and Winogrande. In contrast, Phi-3 Mini, while highly accurate, was the least energy-efficient due to its high power consumption, limiting its suitability for constrained environments. TinyLlama seems ideal for low-power deployments at the cost of accuracy, while Gemma 2 offered a moderate compromise between performance and efficiency.

Our results further highlight the critical role of GPU acceleration, memory bandwidth, and model architecture in minimizing power draw while maintaining performance. Jetson Orin Nano with GPU consistently outperformed CPU-based setups, underscoring the importance of hardware optimization for efficient edge AI deployment. These insights offer practical guidance for selecting SLMs and hardware configurations tailored to specific edge applications. Future work could explore advanced techniques such as adaptive computing, dynamic voltage scaling, and power-aware scheduling. Further gains may be achieved by investigating mixed-precision inference, other quantizations (e.g., GPTQ, AWQ), and model pruning, supporting more sustainable and scalable AI in energy-constrained environments.

\vspace{-0.6mm}
\section{Acknowledgement}
\vspace{-0.5mm}
We thank the anonymous reviewers for their suggestions and
feedback. This research was in part supported by US NSF Grants: SHF-2210744, OAC-2413540, CNS-2244450, AMPS-2229073.

\vspace{-3mm}

\bibliographystyle{IEEEtran}
\bibliography{Ref/ref}

\end{document}